\documentclass{jetpl}
\usepackage{graphicx}
\usepackage{amssymb}
\usepackage{amsmath}
\usepackage{color}
\usepackage{hyperref}
\twocolumn

\newcommand{\DD}{\,\mathrm{d}}

\newcommand{\pd}[2]{\ensuremath{\frac{\partial #1}{\partial #2}}}

\newcommand{\p}{\ensuremath{\hat{\mathbf{p}}}}
\newcommand{\pp}{\ensuremath{\hat{\mathbf{p}}'}}
\newcommand{\mm}{\ensuremath{\mathbf{m}}}
\newcommand{\nn}{\ensuremath{\mathbf{n}}}
\newcommand{\bl}{\ensuremath{\mathbf{l}}}

\title{Spin diffusion in liquid $^3$He confined in planar aerogel}
\rtitle{Spin diffusion in liquid $^3$He confined in planar aerogel}
\sodtitle{Spin diffusion in liquid $^3$He confined in planar aerogel}

\author{V.\,V.\,Dmitriev$^+$\thanks{e-mail: dmitriev@kapitza.ras.ru},\,M.\,S.\,Kutuzov$^\times$,\,L.\,A.\,Melnikovsky$^+$,\,B.\,D.\,Slavov$^{+,*}$,\,A.\,A.\,Soldatov$^{+,*}$,\,A.\,N.\,Yudin$^+$}
\rauthor{V.\,V.\,Dmitriev,\,M.\,S.\,Kutuzov,\,L.\,A.\,Melnikovsky,\,B.\,D.\,Slavov,\,A.\,A.\,Soldatov,\,A.\,N.\,Yudin}
\sodauthor{V.\,V.\,Dmitriev,\,M.\,S.\,Kutuzov,\,L.\,A.\,Melnikovsky,\,B.\,D.\,Slavov,\,A.\,A.\,Soldatov,\,A.\,N.\,Yudin}

\address{$^+$P.\,L. Kapitza Institute for Physical Problems of RAS, 119334 Moscow, Russia}
\address{$^\times$Metallurg Engineering Ltd., 11415 Tallinn, Estonia}
\address{$^*$Moscow Institute of Physics and Technology, 141700 Dolgoprudny, Russia}

\dates{October 30, 2018}{*}

\begin{document}

\abstract{We report the results of theoretical and experimental investigation of spin diffusion in the normal phase of liquid $^3$He confined in planar aerogel: a material consisting of nanostrands which are almost parallel to a specific plane and randomly oriented in this plane. Using spin echo technique we measure the spin diffusion coefficients in the directions perpendicular and parallel to the plane. We see good agreement between the experiment and the theory.}

\maketitle

\section{Introduction}
Measurements of spin diffusion in normal liquid $^3$He confined in high porosity materials (e.g., in aerogels) allow to obtain information about their structure. Aerogel is a rigid system consisting of nanoscale strands. Fermi-liquid quasiparticles in bulk $^3$He have a mean free path $\lambda\propto T^{-2}$ and a corresponding spin diffusion coefficient $D\propto T^{-2}$. At very low temperatures ($T\sim1$\,mK) the density of quasiparticles becomes so small that the aerogel, immersed in liquid $^3$He, limits the free path  and the diffusion is determined by the aerogel structure.

Spin echo technique was used to investigate spin diffusion of liquid $^3$He in different types of aerogel: in isotropic silica aerogels \cite{dsil95,dsil98} and in nematic aerogels \cite{dobn,dnaf} which are nanofabricated materials with almost parallel alignment of strands \cite{nemat}. In the latter case an anisotropic spin diffusion was observed indicating a global anisotropy of nematic aerogel. The spin diffusion along the strands was found to be a few times greater than that in the perpendicular direction. Strong anisotropy of nematic aerogel also leads to existence of a new superfluid phase of $^3$He -- the polar phase \cite{pol}.

\begin{figure}[t]
\centerline{\includegraphics[width=0.9\columnwidth]{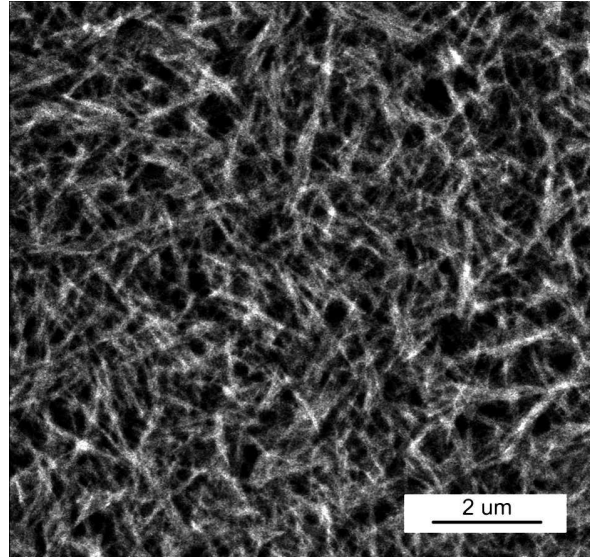}}
\caption{Fig. \thefigure:
Scanning electron microscope image of the free surface of planar aerogel}
\label{sem}
\end{figure}

Here we present results of theoretical and experimental studies of spin diffusion in a different type of anisotropic aerogel-like material, which we call the planar aerogel. Like nematic aerogel, it is a macroscopically uniform system with axial symmetry which consists of long approximately cylindrical strands of nearly the same diameter $d$. The directions of these strands, however, are uniformly distributed in a plane perpendicular to the symmetry axis rather than parallel to it as in nematic aerogel (Fig.~\ref{sem}).

\section{Theory}
The theory of low temperature spin diffusion in normal $^3$He confined in an anisotropic aerogel (particularly in nematic aerogel) was developed in Ref.~\cite{dnaf}. Here we extend it to describe the planar aerogel. At low temperature the influence of collisions between $^3$He quasiparticles can be neglected in comparison with that of aerogel-quasiparticle scattering. Such scattering preserves the quasiparticle energy and (for $^4$He coated strands) spin. At low pressure tangential momentum component is also assumed to be conserved, thus leading to specular quasipartcle reflection \cite{K93}. We will also discuss an opposite limit of diffuse scattering where the tangential momentum components after and before the scattering are independent. This is a natural model for irregular ``rough'' strands.

Let $\DD\sigma^\mm = s^\mm(\p,\pp) \DD\pp$ be the differential $\mathbf{p} \rightarrow \mathbf{p}'$ scattering cross section on a unit surface element of a strand with the outer unit normal \mm. Here and below the hat denotes appropriate unit vector. Then a cylinder of length $L$ has the scattering cross section
\begin{equation*}
\DD\sigma^\nn=
\frac{Ld}{2}
\int\delta(\mm\cdot\nn) \DD\sigma^\mm \DD\mm,
\end{equation*}
where \nn\ is the unit vector along the cylinder axis and $\delta$ is the Dirac delta function. All planar aerogel strands are perpendicular to the unit vector \bl\ directed along the symmetry axis. The scattering cross section per unit volume of such structure is obtained by one more integration:\footnote{One can explicitly verify that
$\int \delta(\mm\cdot\nn) \, \delta(\nn\cdot\bl) \DD\nn=
\left|2\left/\sin \widehat{\mm\bl}\right|\right.$,
where $\widehat{\mm\bl}$ is the angle between \mm\ and \bl.}
\begin{multline*}
\DD\sigma^\bl=
\frac{1}{2\pi}
\int \delta(\nn\cdot\bl) \DD\sigma^\nn \DD\nn=\\
\frac{Ld}{4\pi}
\iint \delta(\mm\cdot\nn)\, \delta(\nn\cdot\bl) \DD\sigma^\mm\DD\nn\DD\mm=\\
\frac{2\nu}{\pi^2 d}
\int \frac{\DD\sigma^\mm}{\left|\sin \widehat{\mm\bl}\right|} \DD\mm
\equiv
\frac{\nu}{\pi d}
S(\p,\pp) \DD\pp
\end{multline*}
where $L=4 \nu/(\pi d^2)$ is the total length of the strands per unit volume, $\nu=1-p$ is the filling factor, $p$ is the aerogel porosity, and
\begin{equation}
\label{s-l}
S(\p,\pp)=
\frac{2}{\pi}
\int \frac{s^\mm(\p,\pp)}{\left|\sin \widehat{\mm\bl}\right|} \DD\mm.
\end{equation}

Let $z$-axis run along \bl. Two distinct principal values of the diffusion tensor are $D^{xx}=D^{yy}$ and $D^{zz}$. The former is obtained from the equation
\begin{equation}
\label{flux}
D^{xx}\, \pd{M}{x} = -\frac{p_F^2}{(2\pi\hbar)^3}
\int
\chi^x(\p)\,
\hat{p}^x \DD\p,
\end{equation}
where $\chi^x(\p)$ is the solution of linearized kinetic equation
\begin{equation}
\label{kinetic-x}
\psi^x \hat{p}^x=
\frac{\nu}{\pi d}
\int
S(\p,\pp)\,
\big(\chi^x(\p)-\chi^x(\pp)\big)
\DD\pp,
\end{equation}
and
\begin{equation*}
\psi^x=\frac{2\pi^2\hbar^3}{p_F m^*} \left(1+F_0^a\right)\pd{M}{x}.
\end{equation*}
Here $M$ is the magnetization, $p_F$ is the Fermi momentum, $m^*$ is the effective mass, and $F_0^a$ is the Fermi-liquid parameter.

To solve such kinetic equation we expand the distribution function $\chi(\p)$ in terms of mutually orthogonal \emph{real} spherical harmonics $Y_{lm} (\p)$:\footnote{Particularly, $m=0$ and $m=1$ harmonics in spherical coordinates are defined as
$Y_{l0}=\sqrt{(2l+1)/(4\pi)} P_l(\cos\theta)$ and
$Y_{l1}=\sqrt{(2l+1)/(2\pi l (l+1))} P_l^1(\cos\theta) \cos\phi$, where $P_l$ and $P_l^1$ are Legendre polynomials.}

\begin{equation*}
\chi(\p)=
\psi \frac{\pi d}{\nu}
\sqrt{\frac{4\pi}{3}}
\sum\limits_{l,m} C^{lm} Y_{lm}(\p).
\end{equation*}
Consider the matrix $\mathsf{S}$ of the collision operator in the right-hand side of Eq.~\eqref{kinetic-x} defined by the expression
\begin{equation}
\label{collision-matrix}
\int
Y_{l_1 m_1}(\p)\,
 S(\p,\pp)\,
\big(Y_{l_2 m_2}(\p)-Y_{l_2 m_2}(\pp)\big)
\DD\pp\DD\p.
\end{equation}
Due to the symmetry of the system under parity inversion, the matrix elements $l_1\not \equiv l_2 \pmod 2$ are identically zero. Another selection rule is a consequence of combined full $z$-axial and vertical mirror plane symmetry: $m_1=m_2$.

The left hand side of Eq.~\eqref{kinetic-x} is, up to a factor, $Y_{11}$. The solution of this equation is therefore
\begin{equation*}
\chi^x (\p)=
\psi^x \frac{\pi d}{\nu}
\sqrt{\frac{4\pi}{3}}
\sum\limits_{l=1,3,5,\dots}^{\infty} U^l Y_{l1}(\p),
\end{equation*}
where $U^l$ is the solution of the matrix equation
\begin{equation}
\label{matrixeq}
(1\ 0\ 0\ \dots)^T=
\mathsf{S} \mathbf{U}.
\end{equation}
The diffusion coefficient depends, in fact, (see Eq.~\eqref{flux}) only on the first component of the solution $U^1$:
\begin{equation*}
D^{xx}=
U^1
\left(1+F_0^a\right)
\frac{\pi v_F d}{3\nu}.
\end{equation*}

The kinetic equation solution for the diffusion along $z$-axis
\begin{equation}
\label{kinetic-z}
\psi^z \hat{p}^z=
\frac{\nu}{\pi d}
\int
S(\p,\pp)\,
\big(\chi^z(\p)-\chi^z(\pp)\big)
\DD\pp
\end{equation}
is sought for in the form
\begin{equation*}
\chi^z (\p)=
\psi^z \frac{\pi d}{\nu}
\sqrt{\frac{4\pi}{3}}
\sum\limits_{l=1,3,5,\dots}^{\infty} V^l Y_{l0}(\p).
\end{equation*}
The diffusion coefficient in this case is given by
\begin{equation*}
D^{zz}=
V^1
\left(1+F_0^a\right)
\frac{\pi v_F d}{3\nu}.
\end{equation*}

In the case of specular scattering (denoted by the subscript ``S'' below), elementary cross section~\cite{dnaf} is

\begin{equation*}
s_\text{S}^\mm(\p,\pp)=-(\p \mm)
\begin{cases}
0,								&(\mm \p)>0;\\
\delta\left(
    \pp-\p+2\mm (\p \mm)
    \right)								&\text{otherwise.}
\end{cases}
\end{equation*}
This expression can be simplified
\begin{equation*}
s_\text{S}^\mm(\p,\pp)=\frac{1}{4}
\delta\left(\mm - \frac{\pp-\p}{\left|\pp-\p\right|}\right)
\end{equation*}
and plugged into Eq.~\eqref{s-l}:
\begin{multline}
\label{s-l-specular}
S_{\text{S}}(\p,\pp)=
\frac{1}{2\pi}
\frac{1}{\left|\sin \angle \left(\p-\pp, \bl\right)\right|}=\\
\frac{1}{\pi \sqrt{2} }
\sqrt{\frac{{1-\cos\theta\cos\theta'-\sin\theta\sin\theta'\cos\Delta}}{{\sin^2\theta+\sin^2\theta'-2\sin\theta\sin\theta'\cos\Delta}}},
\end{multline}
where the spherical angles $(\theta,\phi)$ and $(\theta',\phi')$ correspond to \p\ and \pp\ and $\Delta=\phi-\phi'$.

To calculate the collision operator matrix \eqref{collision-matrix},
the cross section \eqref{s-l-specular} must be integrated with respect to $\theta$, $\phi$, $\theta'$, and $\phi'$. One integration can be saved using the above mentioned symmetry.\footnote{Namely, if a function $h$ depends on $\left|\phi-\phi'\right|$ but not on individual angles $\phi$ and $\phi'$, then $\int h(\phi-\phi') \DD\phi\DD\phi' = 2\pi \int h(\phi) \DD\phi$ and $\int h(\phi-\phi') \cos\phi\, \cos \phi'\DD\phi\DD\phi' = \pi \int h(\phi) \cos\phi \DD\phi$.} Remaining triple integrals are evaluated numerically to solve Eq.~\eqref{matrixeq} with the help of SciPy \cite{scipy} and particularly its interface to \texttt{QUADPACK} library \cite{quadpack}. It turns out that sufficient precision is reached already for the matrices as small as $10\times 10$. This procedure gives $U_\text{S}^1 = 0.4249$, $V_\text{S}^1 = 0.2153$ and
\begin{align}
D_\text{S}^{xx} &=0.445 \left(1+F_0^a\right) \frac{v_F d}{\nu}, \label{D-S-xx}\\
D_\text{S}^{zz} &=0.226 \left(1+F_0^a\right) \frac{v_F d}{\nu}. \label{D-S-zz}
\end{align}
It is worth noting that $D_\text{S}^{zz}$ is close to the lateral component of the diffusion tensor (spin flux is also perpendicular to the strands)
\begin{equation*}
D_\text{S}^\perp=
\frac{3\pi^2}{128}
\left(1+F_0^a\right) \frac{v_F d}{\nu}
\approx
0.231
\left(1+F_0^a\right) \frac{v_F d}{\nu}
\end{equation*}
found for specular scattering in nematic aerogel \cite{dnaf}.

In the diffuse scattering model (denoted by the subscript ``D'') we failed to find an analytical expression for the scattering cross-section of planar aerogel $S_\text{D}(\p,\pp)$ by substituting the elementary diffuse cross-section directly into Eq.~\eqref{s-l}, as in Eq.~\eqref{s-l-specular}. This problem can be solved by the following trick. As explained earlier, diffusion depends only on diagonal $m_1=m_2$ elements of the matrix $\mathsf{S}$ (other elements are identically zero). The $m=0$ elements for the planar aerogel, formed by the strands uniformly distributed in $xy$-plane, coincide with the same elements of a collision operator matrix computed for a nematic aerogel whose strands are aligned with an arbitrary vector \nn\ in $xy$-plane.
Similarly, $m=1$ elements are calculated using the identity:
\begin{multline*}
\int h(\phi-\alpha\, ,\, \phi'-\alpha) \cos\phi\, \cos \phi' \, \frac{\DD\alpha}{2\pi}\DD\phi\DD\phi' =\\
\frac{1}{2}\int h(\phi,\phi') \cos(\phi-\phi') \DD\phi \DD\phi'.
\end{multline*}
We employ nematic aerogel scattering cross-section
$\DD\sigma_\text{D}^\nn\equiv
\nu S_\text{D}^\nn (\p,\pp) \DD\pp / (\pi d)$,
where \cite{dnaf}
\begin{equation}\label{nematic_diffuse}
S_\text{D}^\nn (\p,\pp)=
\frac{1}{\pi}\sin\tilde\theta \, \sin\tilde\theta' \,
\bigl|
\sin \tilde\Delta -
\tilde\Delta \cos \tilde\Delta
\bigr|
\end{equation}
and the spherical angles $\tilde\theta$, $\tilde\theta'$, and $\tilde\Delta=\tilde\phi-\tilde\phi'$ are relative to \nn. If \nn\ is selected in $x$ direction, then
\begin{equation*}
\cos\tilde\theta=\sin\theta\cos\phi,
\quad
\cos\tilde\theta'=\sin\theta' \cos\phi',
\end{equation*}
\begin{equation*}
\cos\tilde\Delta=
\frac{\cos\theta \cos\theta' + \sin\theta \sin\theta' \sin\phi \sin\phi'}{\sin\tilde\theta \sin\tilde\theta'}.
\end{equation*}
Genuine quadruple integrals with respect to $\theta$, $\phi$, $\theta'$, and $\phi'$ must be evaluated when Eq.~\eqref{nematic_diffuse} is substituted in Eq.~\eqref{collision-matrix}. Significant speed-up is achieved by implementing the integrand in C language. Computations give $U_\text{D}^1 = 0.2529$, $V_\text{D}^1 = 0.1788$ and
\begin{align}
D_\text{D}^{xx} &=0.265 \left(1+F_0^a\right) \frac{v_F d}{\nu}, \label{D-D-xx}\\
D_\text{D}^{zz} &=0.187 \left(1+F_0^a\right) \frac{v_F d}{\nu}. \label{D-D-zz}
\end{align}
It is again instructive to compare $D_\text{D}^{zz}$ with
\begin{equation*}
D_\text{D}^\perp =
\frac{\pi^2}{2\pi^2+32}
\left(1+F_0^a\right)
\frac{ v_F d}{\nu}
\approx
0.191
\left(1+F_0^a\right)
\frac{ v_F d}{\nu},
\end{equation*}
obtained in Ref.~\cite{dnaf} for spin diffusion across nematic aerogel in diffuse scattering model.

\section{Details of experiment}
The sample of planar aerogel used in our experiments was produced from an aluminum silicate (mullite) nematic aerogel consisting of strands with diameters $d\approx10$\,nm. The process includes dispersion of the aerogel in alcohol so the strands break up and detach from one another. Then the mixture is dried at room temperature and heat treated (baked), to generate some bonding between contacting fibers. The resulting structure is a fibrous network mostly oriented in one plane (Fig.~\ref{sem}) with porosity of about 88\% and the overall density $\rho\approx350$\,mg/cm$^3$ (the density of mullite $\rho_0\approx3$\,g/cm$^3$). From Fig.~\ref{sem} it is seen that lengths of separate strands are of the order of 1\,$\mu$m. Similar structures are used for fabrication of filtration membranes and described in Refs.~\cite{kut1,kut2}.

Three plates with sizes $4\times4$\,mm were cut from the original sample (having the form of a disk with a thickness of $\approx1$\,mm), stacked on top of each other, and placed in a separate cell of our experimental chamber. The experimental chamber (made of Stycast-1266 epoxy resin) was similar to that described in Ref.~\cite{cell}.

Experiments were carried out using spin echo technique in the magnetic field of 140\,Oe (corresponding NMR frequency is 453\,kHz) at the pressure of 2.9\,bar. In order to avoid a paramagnetic signal from solid $^3$He on the surface of aerogel strands, the sample was coated by $\approx2.5$ atomic layers of $^4$He before filling the chamber with $^3$He. The external static magnetic field was perpendicular to the planar aerogel plane. Two systems of gradient coils were used to apply the field gradient in directions parallel and perpendicular to the plane. Necessary temperatures were obtained by a nuclear demagnetization cryostat and measured by a quartz tuning fork. The temperature was determined in assumption that the resonance linewidth of the fork in normal $^3$He is inversely proportional to the temperature \cite{fork}. At high temperatures the diffusion coefficient in aerogel should be the same as in bulk $^3$He (we used the data from Ref.~\cite{diff0}) that allowed us to calibrate the temperature scale.

We obtained spin echo decay curves by measuring the amplitude of the echo after $\pi/2-\tau-\pi$ pulses, where $\tau$ is the delay between pulses. The measurements were carried out at temperatures 1.5--80\,mK for two directions of magnetic field gradient (parallel and perpendicular to the plane of planar aerogel) and at several values of the gradients (265--786\,mOe/cm).

\section{Experimental results}
The spin echo amplitude is obtained from Bloch-Torrey equations \cite{Tor} and given by
\begin{equation}\label{ampl}
I=I_0\exp(-2\tau/T_2-A\tau^3),
\end{equation}
where $T_2$ is a spin-spin relaxation time and $A$ for an anisotropic media has a form of
\begin{equation}
A=\frac{2}{3}\gamma^2D^{lm}G^lG^m.
\end{equation}
Here $\gamma$ is a gyromagnetic ratio of $^3$He, $\bf G$ is a gradient vector of the external magnetic field.

\begin{figure}[t]
\center
\includegraphics[width=\columnwidth]{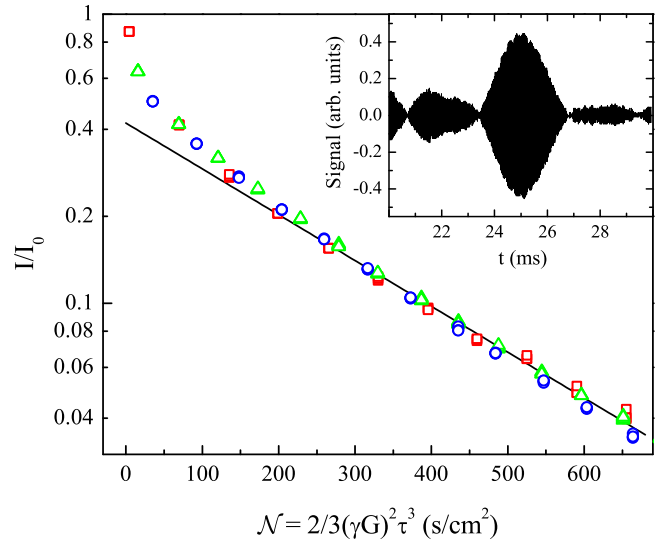}
\caption{Fig. \thefigure:
Spin echo amplitude (normalized to the amplitude at $\tau=0$) versus $\mathcal{N}=\frac{2}{3}(\gamma G)^2\tau^3$ for different field gradients applied in $z$ direction. $G=765$\,mOe/cm (squares), 515\,mOe/cm (triangles), 265\,mOe/cm (circles); $T\approx1.6$\,mK. Solid line is the best fit to the data at $\mathcal{N}>200$\,s/cm$^2$ by Eq.~(\ref{ampl}). Inset: typical echo signal. $G=765$\,mOe/cm, $T\approx1.6$\,mK, $\tau=12.5$\,ms}
\label{echo}
\end{figure}
\begin{figure}[t]
\center
\includegraphics[width=\columnwidth]{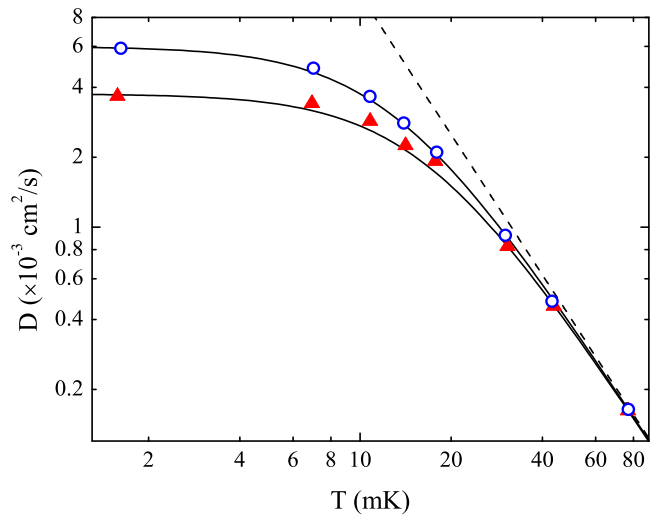}
\caption{Fig. \thefigure:
Temperature dependence of the spin diffusion coefficients in planar aerogel: $D^{xx}(T)$ (circles) and $D^{zz}(T)$ (triangles)}
\label{diff}
\end{figure}

A typical echo signal of $^3$He in the planar aerogel is shown in the inset of Fig.~\ref{echo}. The value of spin diffusion coefficient at fixed temperature $T$ is determined from echo amplitudes measured at different $\tau$ after fitting to Eq.~(\ref{ampl}). The observed dependence of the echo amplitude $I/I_0$ on $G^2\tau^3$ does not depend on field gradient at all used temperatures, so the term with $T_2$ in Eq.~(\ref{ampl}) can be neglected. An example data set is shown in Fig.~\ref{echo}. We note that at $\mathcal{N}\le200$\,s/cm$^2$ experimental points deviate from the linear dependence due to the presence of bulk $^3$He in the cell filling tube and in the gaps between the aerogel sample and the cell walls. At low temperatures ($T\lesssim40$\,mK) the spin diffusion in bulk $^3$He is much greater than that in aerogel. Therefore, the relative contribution of bulk $^3$He into the total echo signal rapidly decreases with the increase of $\tau$, and spin diffusion coefficients in aerogel can be determined from the data where the linear law is observed ($\mathcal{N}>200$\,s/cm$^2$ for the data in Fig.~\ref{echo}).

Temperature dependencies of the spin diffusion coefficients shown in Fig.~\ref{diff} were measured for two orientations of the gradient: parallel ($D^{xx}$) and perpendicular ($D^{zz}$) to $xy$-plane. Each set of data is fitted by the following equation:
\begin{equation}\label{diffsum}
D^{-1}(T)=D_{bulk}^{-1}(T)+D^{-1},
\end{equation}
where the contributions of collisions between quasiparticles $D_{bulk}\propto T^{-2}$ (the diffusion coefficient in bulk $^3$He) and that of quasiparticle-aerogel scattering $D\equiv D(0)$ are separated. Solid lines in the graph are the best fits to Eq.~(\ref{diffsum}), the dashed line is the diffusion coefficient in bulk $^3$He (obtained from extrapolation to $P=2.9$\,bar of the experimental data in Ref.~\cite{diff0}). Thus, we get principal values of the spin diffusion tensor in planar aerogel in zero temperature limit: $D^{xx}=0.0059$\,cm$^2$/s, $D^{zz}=0.0036$\,cm$^2$/s. The accuracy of these values is estimated as $\pm10$\%.

\section{Discussion}
We define zero-temperature effective mean free paths $\lambda_{z}$ and $\lambda_{x}$ of $^3$He quasiparticles in planar aerogel by the equation \cite{dnaf}: $D=v_F\lambda\left(1+F_0^a\right)/3$. For $v_F=5397$\,cm/s and $F_0^a=-0.717$ \cite{3hecalc} we get $\lambda_{z}=71$\,nm and $\lambda_{x}=116$\,nm.

For our sample of planar aerogel $d\approx10$\,nm and $\nu=\rho/\rho_0\approx0.117$. Thus, from Eqs.~(\ref{D-S-xx},\ref{D-S-zz},\ref{D-D-xx},\ref{D-D-zz}) we expect to have the following spin diffusion coefficients for specular scattering $D^{xx}_\text{S}=0.00583$\,cm$^2$/s, $D^{zz}_\text{S}=0.00296$\,cm$^2$/s and for diffuse scattering $D^{xx}_\text{D}=0.00613$\,cm$^2$/s, $D^{zz}_\text{D}=0.00245$\,cm$^2$/s.
At 2.9\,bar the specular scattering is expected. We note that inaccuracies in $d$ and $\nu$ do not influence the ratio $D^{xx}/D^{zz}$, and the discrepancy between the experimentally observed $D^{xx}/D^{zz}=1.64$ and
$D^{xx}_\text{S}/D^{zz}_\text{S}=1.97$ is probably due to incomplete alignment of aerogel strands in one plane. It is worth to mark that for diffuse scattering the theory predicts $D^{xx}_\text{D}/D^{zz}_\text{D}=1.41$.

The observed strong anisotropy of $^3$He spin diffusion is of a particular interest for NMR experiments with superfluid $^3$He in planar aerogel where the A phase with the orbital vector oriented perpendicular to the plane is expected to emerge \cite{LIM} as well as the effect of a magnetic scattering can be manifested, which was presumably the case for the superfluid $^3$He in nematic aerogel \cite{mag}.

M.S.K. prepared the aerogel sample. L.A.M. and B.D.S. developed the theory and were supported in part by the Basic Research Program of the Presidium of Russian Academy of Sciences ``Actual problems of low temperature physics''. The experiments were carried out by V.V.D., A.A.S., and A.N.Y. and were supported by grant of the Russian Science Foundation (project \#\,18-12-00384).

\end{document}